\documentclass{iau} \usepackage{graphicx}

\title[The ISM content in ALMA deep fields] 
{The interstellar medium content of galaxies in the ALMA era}

\author[Manuel Aravena] {Manuel Aravena$^1$} \affiliation{$^1$N\'ucleo de Astronom\'{\i}a, Facultad de Ingenier\'{\i}a y Ciencias, \\ Universidad Diego Portales, 8370191 Santiago,
Chile \\ email: {\tt manuel.aravenaa@mail.udp.cl}}

\pubyear{2019} \volume{352}  
\jname{Uncovering early galaxy evolution in the ALMA and JWST era} \editors{E. da Cunha, J.
Hodge, J. Afonso \& L. Pentericci, eds.} 
\begin{document}

\maketitle

\begin{abstract} The advent of the Atacama Large Millimeter/submillimeter Array (ALMA) has enabled a new era for studies of the formation and assembly of distant galaxies.
Cosmological deep field surveys with ALMA and other interferometers have flourished in the last few years covering wide ranges of galaxy properties and redshift, and allowing us
to gain critical insights into the physical mechanisms behind the galaxy growth. Here, we present a brief review of recent studies that aim to characterize the interstellar
medium properties of galaxies at high redshift ($z>1$), focusing on blank-field ALMA surveys of dust continuum and molecular line emission. In particular, we show
recent results from the ALMA Spectroscopic Survey in the Hubble Ultra Deep Field (ASPECS) large program. 

\keywords{galaxies: high-redshift, galaxies: evolution, galaxies: ISM}
\end{abstract}

\firstsection 
\section{Introduction}

\begin{figure*}
\centering
\includegraphics[scale=0.3]{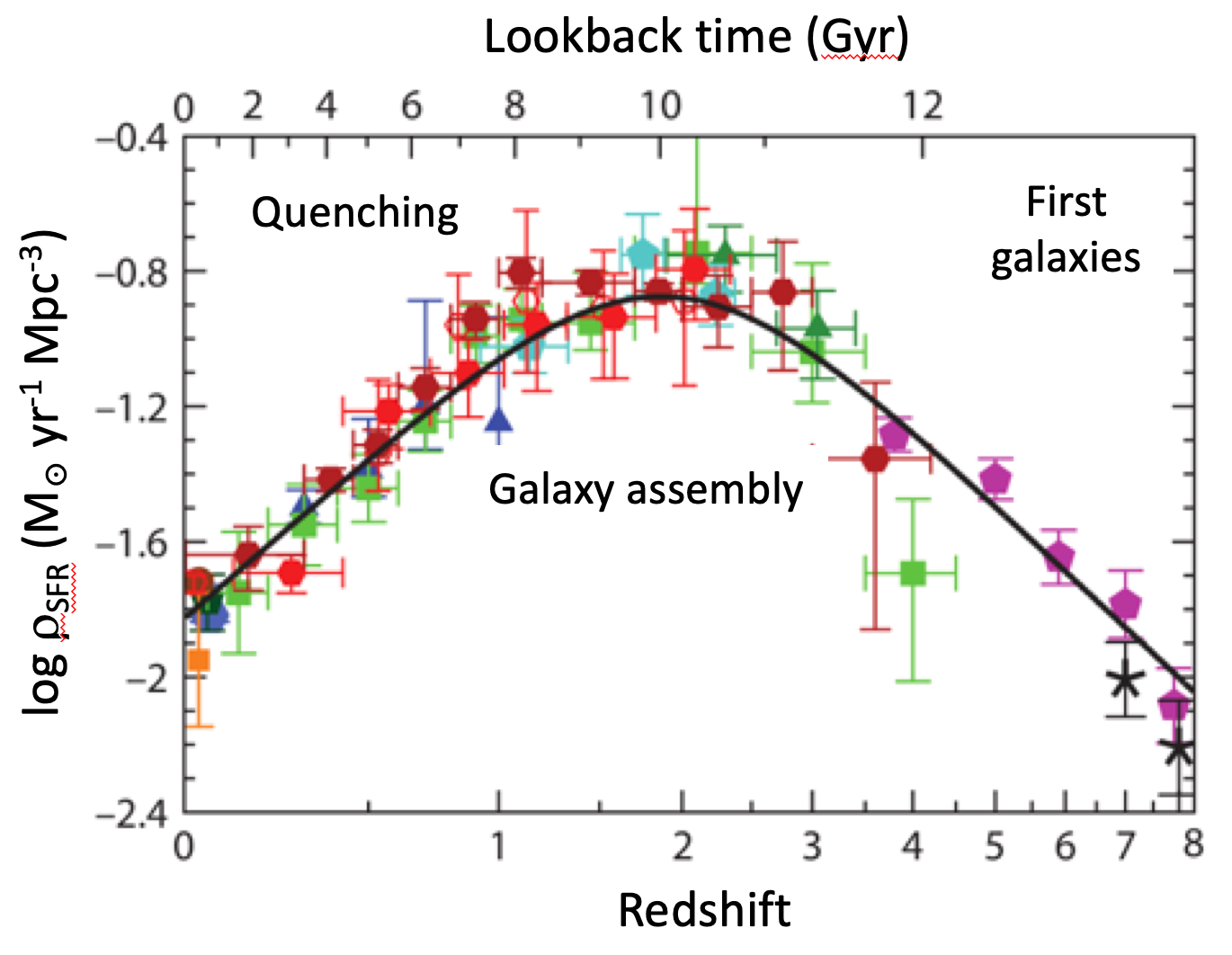}
\includegraphics[scale=0.3]{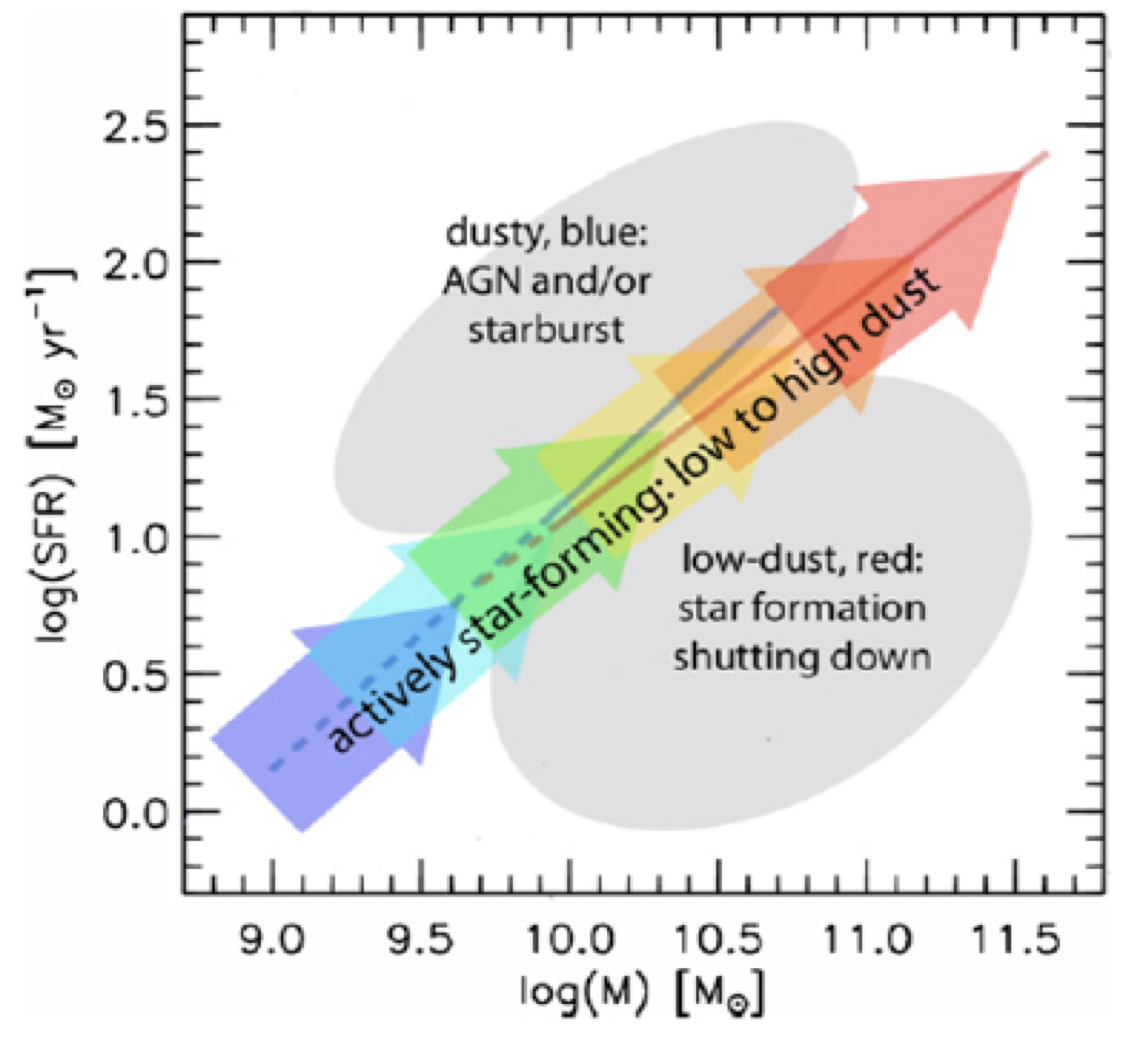}
\caption{{\it (Left:)} Evolution of the cosmic SFR density with redshift adapted from  Madau \& Dickinson (2004). 
The main epochs for galaxy formation are highlighted. While the rise and fall of star formation with redshift has been known for 
several years, little is known about the reasons behind this evolution. {\it (Right:)} Cartoon view of the location of MS galaxies, starburst and quenched galaxies 
in the SFR vs stellar mass plot (Whitaker et al. 2014). Targeted studies have so far focused in follow-up observations of highly star forming galaxies in top part of this diagram.\label{fig:csfrd_ms}}
\end{figure*}

One of the most outstanding challenges in galaxy evolution today is to understand how galaxies obtain their cold gas to sustain star formation activity through cosmic time. A
significant advance in the last 20 years has been the determination of the cosmic star formation rate (SFR) density out to the end of the Epoch of Reionization ($z\sim6-8$). From
this epoch on, the cosmic SFR density rose gradually up to a peak level ($1<z<3$), before smoothly declining by an order of magnitude towards the present day ($0<z<1$). The period
of peak star-formation constitutes the main epoch of galaxy assembly, when roughly half the stars in the Universe were formed (\cite[Fig. \ref{fig:csfrd_ms}; Madau \& Dickinson, 2014]{madau14}). However, why the
cosmic SFR density follows this evolution is a matter of increasingly active investigation, with various studies aiming to measure the evolution of the cosmic density of molecular
gas, as the latter is the fuel for active star formation. Critical questions are related to whether there are fundamental variations in average molecular gas content in galaxies
through cosmic time that would produce the observed shape in the cosmic SFR density, or whether this is produced by changes in the molecular gas efficiency (e.g., \cite[Decarli et al. 2014]{decarli14}).

A key finding has been that since early times, $z<6$, most galaxies display a linear relation between their stellar masses and SFRs, forming what is usually referred to as the main
sequence of star-forming galaxies (Fig. \ref{fig:csfrd_ms}). Galaxies with SFRs above and below this sequence are respectively termed ‘starburst’, and ‘quiescent’ (\cite[e.g., Elbaz et al. 2007]{elbaz07}). Most
main sequence galaxies resemble clumpy rotating disks, whereas starburst galaxies are typically associated to galaxy collisions and mergers (\cite[Daddi et al. 2010; Tacconi et al. 2013]{daddi10, tacconi13}). 
Main sequence galaxies constitute the population that dominates the cosmic SFR density, and thus they are critical to understand the transformation of galaxies through
cosmic time (e.g. \cite[Magnelli et al. 2012]{magnelli12}).

Efforts to measure the molecular gas content in galaxies at high-redshift have focused on detection of the $^{12}$CO line and dust continuum emission, as these are the most direct
tracers of the total molecular gas (\cite[e.g., Carilli \& Walter et al. 2013]{carilli13}). Dust continuum emission has received particular interest since it is easier to detect observationally and benefits from a negative K-correction
out to high redshifts (\cite[e.g. Scoville et al. 2014]{scoville14}). These tracers, however, are not free from uncertainties, since CO and dust measurements rely on assumptions on the CO-to-gas mass conversion factor and the
gas-to-dust ratio, respectively, which are dependent on metallicity and environment and thus can vary by factors of a few (even though they can be calibrated under certain
conditions). Recent studies have proposed the use of [CI] and even [CII] line emission as tracers for molecular gas mass (\cite[Valentino et al. 2018; Zanella et al. 2018]{valentino18, zanella18}). However, similarly to CO and dust, [CI] determinations
require assumptions on the [CI]/H$_2$ abundance ratio and the use of [CII] as a molecular gas tracer is not so direct.

\section{Approaches to detect the ISM at high-z}

Significant efforts have been devoted in the last 20 years to conduct blank-field sub-millimeter surveys over large areas of the sky with bolometer cameras in single dish
telescopes (\cite[e.g. Smail et al. 1997; Hughes et al. 1998; Weiss et al. 2009]{smail97,hughes98,weiss09}). These surveys discovered an important population of dust-rich starburst 
galaxies at high redshift that contributed a significant fraction of the cosmic SFR density (\cite[e.g. Casey et al. 2014]{casey04}).
These sources, called ``submillimeter galaxies'' (SMGs), turned out to be inconspicuous at optical wavelengths due to heavy dust obscuration. Although detailed characterization
has been difficult, it is relatively clear that SMGs are massive galaxies undergoing episodes of enhanced star formation, typically catalogued as starburst galaxies in the stellar
mass versus SFR diagram, with SFRs above 200 $M_\odot$ yr$^{-1}$ and redshifts in the range $1-4$.

Since these (sub)millimeter surveys are only able to grasp the brightest dusty star-forming galaxies in the sky, pinpointing the typical population of galaxies at $z\sim1-3$ with typical
SFRs $\sim50-200\ M_\odot$ yr$^{-1}$, has required targeted observations of dust continuum and molecular gas emission in galaxies that have been pre-selected through their optical
colors and/or faint IR emission detected in {\it Herschel} IR surveys (\cite[e.g. Daddi et al. 2008, 2010; Dannerbauer et al. 2009; Aravena et al. 2010, 2012; Saintonge et al. 2013;
Tacconi et al. 2010, 2013, 2018;  Bolatto et al. 2015; Scoville et al. 2017; Freundlich et al. 2018]{daddi08, daddi10, dannerbauer09, aravena10, aravena12, saintonge13,
taconni10, tacconi13, tacconi18, bolatto15, scoville17, freundlich18}). These follow-up programs have been extremely successful to measure molecular gas masses
 of a large number of galaxies, enabling the determination of observational scaling relations between various interstellar medium (ISM) parameters (gas fractions, efficiencies, distance to the MS, etc) and 
 the establishment of a framework to understand the evolution of galaxies (\cite[e.g. Genzel et al. 2015; Tacconi et al. 2018]{genzel15, tacconi18}).

A complementary approach to reach the fainter dusty galaxies has been the use of archival data toward an increasing number of extragalactic pointings and calibrator fields observed
with the Atacama Large Millimeter/submillimeter Array (ALMA) (\cite[Oteo et al. 2015, 2017; Carniani et al. 2014; Fujimoto et al. 2016, 2017]{oteo15, oteo17, carniani14,
fujimoto16, fujimoto17}). The main motivation of these projects is to take advantage of the already available data to look for
galaxies in the field of the main target of the ALMA observations, which due to the large number of visits of those fields can reach great depths. 
These surveys have been successful in putting constraints on the (sub)millimeter number count distribution,
however, they typically do not have enough multi-wavelength ancillary data to measure the galaxies' properties.

\begin{figure*}
\centering
\includegraphics[scale=0.38]{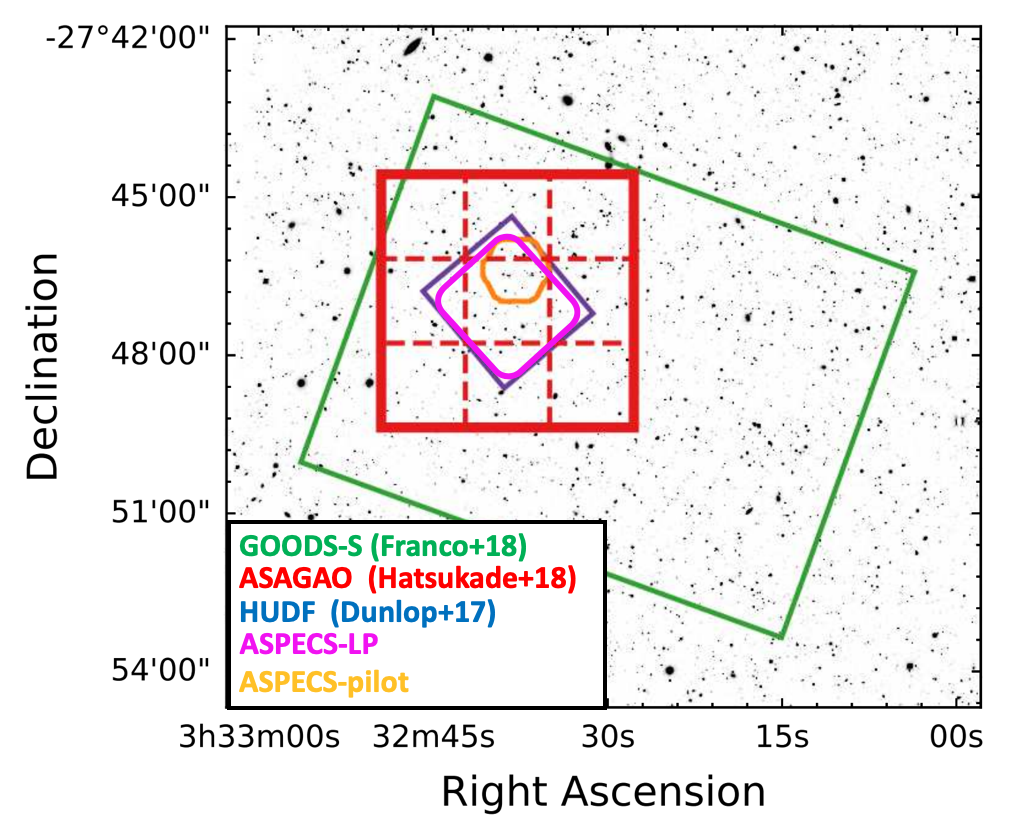}
\includegraphics[scale=0.38]{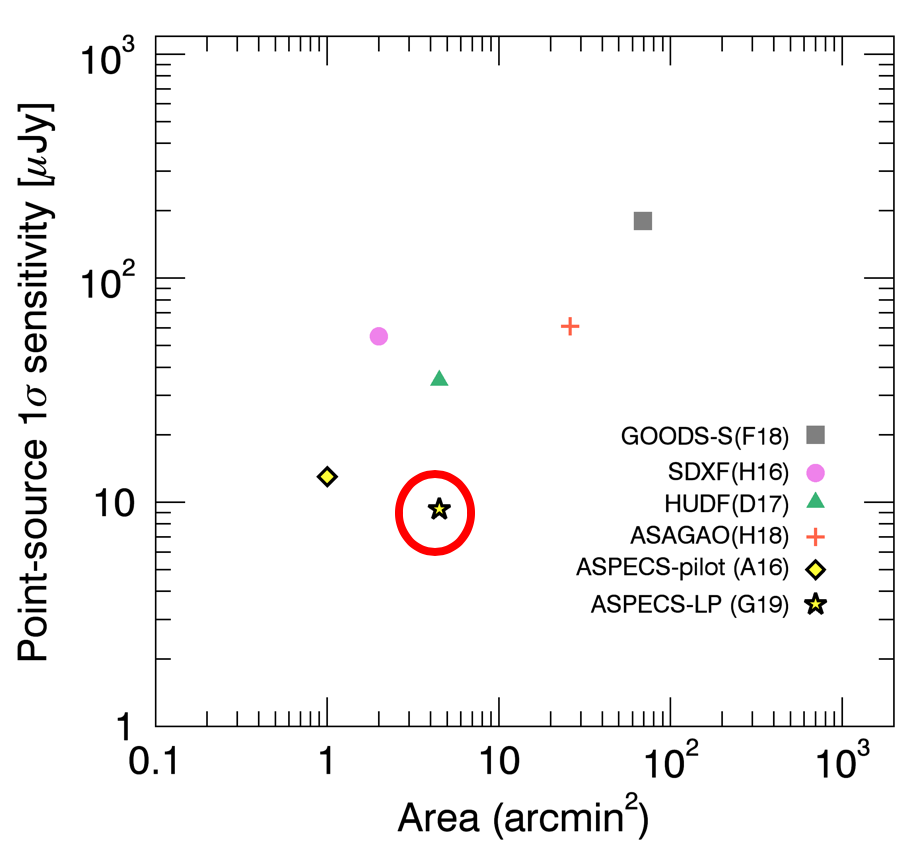}
\caption{{\it (Left:)} Relative location of the different ALMA millimeter deep surveys in the GOODS-S field (adapted from Hatsukade et al. (2018). {\it (Right:)}
Comparison of the area and point-source sensitivities achieved by the various ALMA deep field efforts at 1mm. Current surveys have been following a ``wedding cake'' approach.
The red circle highlights the great sensitivity and area covered by the ASPECS large program.}
\label{fig:layouts}
\end{figure*}

\section{Millimeter continuum deep fields}

Recent deep millimeter continuum surveys over increasingly large contiguous areas with ALMA toward cosmological deep fields are revolutionizing galaxy evolution studies by
accessing the dust emission from the faint star-forming galaxy population.

These surveys have been roughly following a ``wedding cake'' approach, covering large areas at shallower depths and narrow areas deeper, and concentrating in the 1-mm band due to the efficiency to reach good depth and area coverage, while targeting a sufficiently
high part of the Rayleigh-Jeans tail of the dust spectral energy distribution of galaxies at $z>1$. Figure \ref{fig:layouts} compares the depth and area achieved by the ALMA millimeter continuum surveys to date. In order of decreasing area,
current  millimeter continuum surveys include: The ALMA survey of the GOODS-S field (\cite[Franco et al. 2018]{franco18}), the ALMA twenty-Six Arcmin$^2$ Survey of GOODS-S at One-millimeter
(\cite[ASAGAO; Hatsukade et al. 2018]{hatsukade18}), the ALMA deep field in SSA22 (\cite[Umehata et al. 2018]{umehata18}), the ALMA 1.3-mm continuum survey of the Hubble Ultra Deep Field (\cite[HUDF; Dunlop et al. 2017]{dunlop17}) and
The ALMA Subaru-XMM Deep field (SXDF) ALMA 2-arcmin$^2$ deep field survey (\cite[Hatsukade et al. 2016]{hatsukade16}).

The availability of ancillary data has been key to provide a detailed characterization of the identified galaxies. For example, the galaxies detected by the ALMA 1.3-mm survey of
the HUDF, with $S_{\rm 1.3mm}>120\mu$Jy ($3.5\sigma$), are shown to be massive galaxies mostly located in the main-sequence of star formation with stellar masses $M_{\rm
stars}>2\times10^{10}\ M_\odot$ and typical redshifts $z\sim2$ (\cite[Dunlop et al. 2017]{dunlop17}). Similar results are found in other ALMA surveys (\cite[Hatsukade et al. 2016]{hatsukade16}). The sample of galaxies drawn from these ALMA surveys can thus be seen as galaxies as massive as SMGs, but that are not necessarily undergoing a starburst phase with enhanced SFR.
This is exemplified by the $5-10\times$ lower 1-mm fluxes observed. Figure \ref{fig:layouts} shows the location of the various ALMA 1mm deep field surveys 
so far conducted in the GOODS-S field, and compares the area and point source sensitivities achieved so far.

\begin{figure*}
\centering
\includegraphics[scale=0.35]{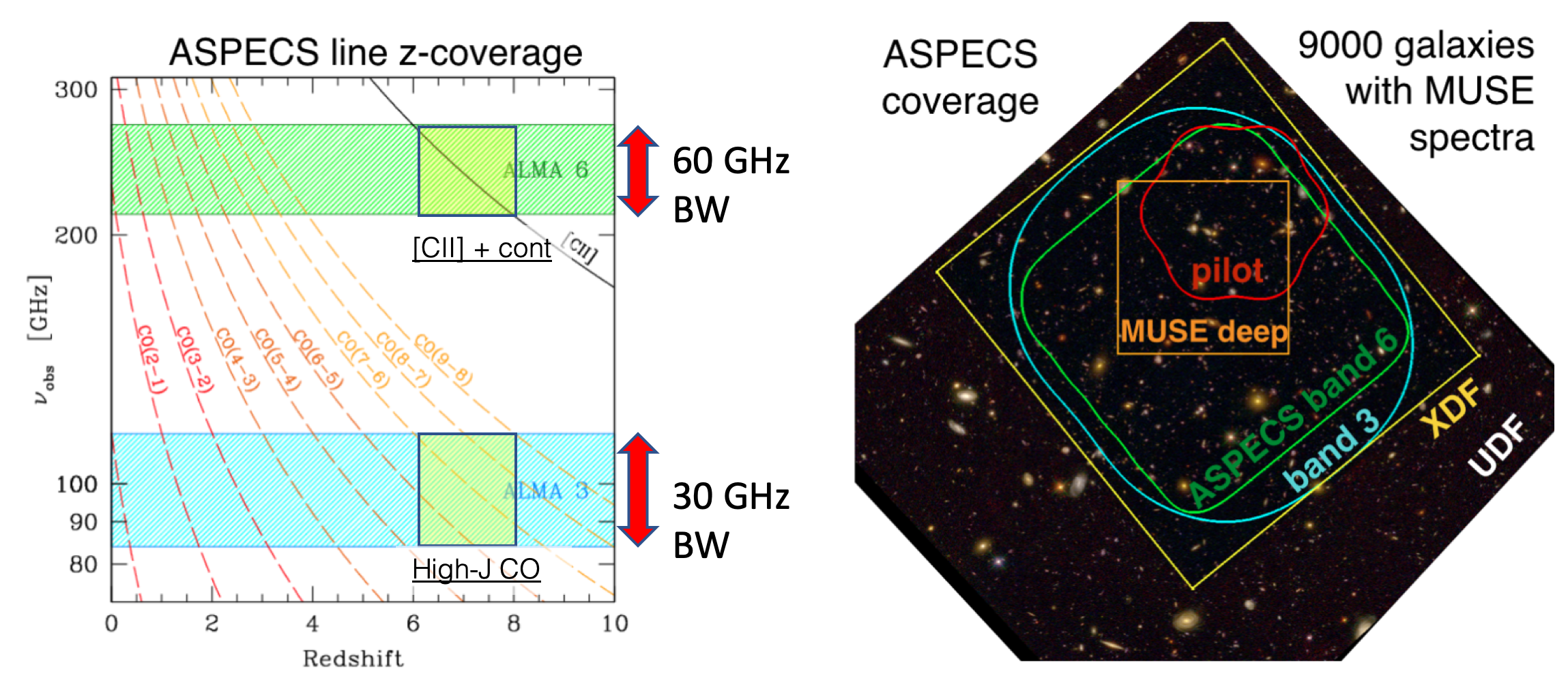}
\caption{{\it (Left:)} Frequency and redshift coverage for CO and [CII] line emission for the ASPECS program. The cyan and green bands highlight the frequency
coverage in ALMA bands 3 and 6, representing bandwidths (BW) of 30 and 60 GHz, respectively. The yellow boxes represent the coverage for CO/[CII] at $z=6-8$.
{\it (Right:)} Sky coverage of the ASPECS pilot and large programs with respect to the location of the HUDF, XDF and MUSE coverage in the field.} 
\label{fig:aspecs_setup}
\end{figure*}

\section{The ALMA Spectroscopic Survey in the HUDF}

The ALMA Spectroscopic Survey in the HUDF (ASPECS) project pioneered a parallel, complementary approach to pure continuum millimeter deep field observations, following previous
results using the Plateau de Bureau Interferometer (PdBI; Decarli et al. 2014; Walter et al. 2014) and the Very Large Array (VLA; Aravena et al. 2012). By scanning the full ALMA
bands 3 and 6 over the ranges 84-115 GHz and 212-275 GHz, the ASPECS observations are able to access the CO and [CII] emission lines in the redshift ranges $z=0-6$ and $6-8$,
respectively. By collapsing along the frequency axis, great sensitivity in the 3mm and 1mm continuum can be achieved. Figure \ref{fig:aspecs_setup} shows the ASPECS 
survey design, including the frequency coverage and location with respect to the HUDF area.

\subsection{ASPECS pilot} 
\begin{figure*}
\centering
\includegraphics[scale=0.6]{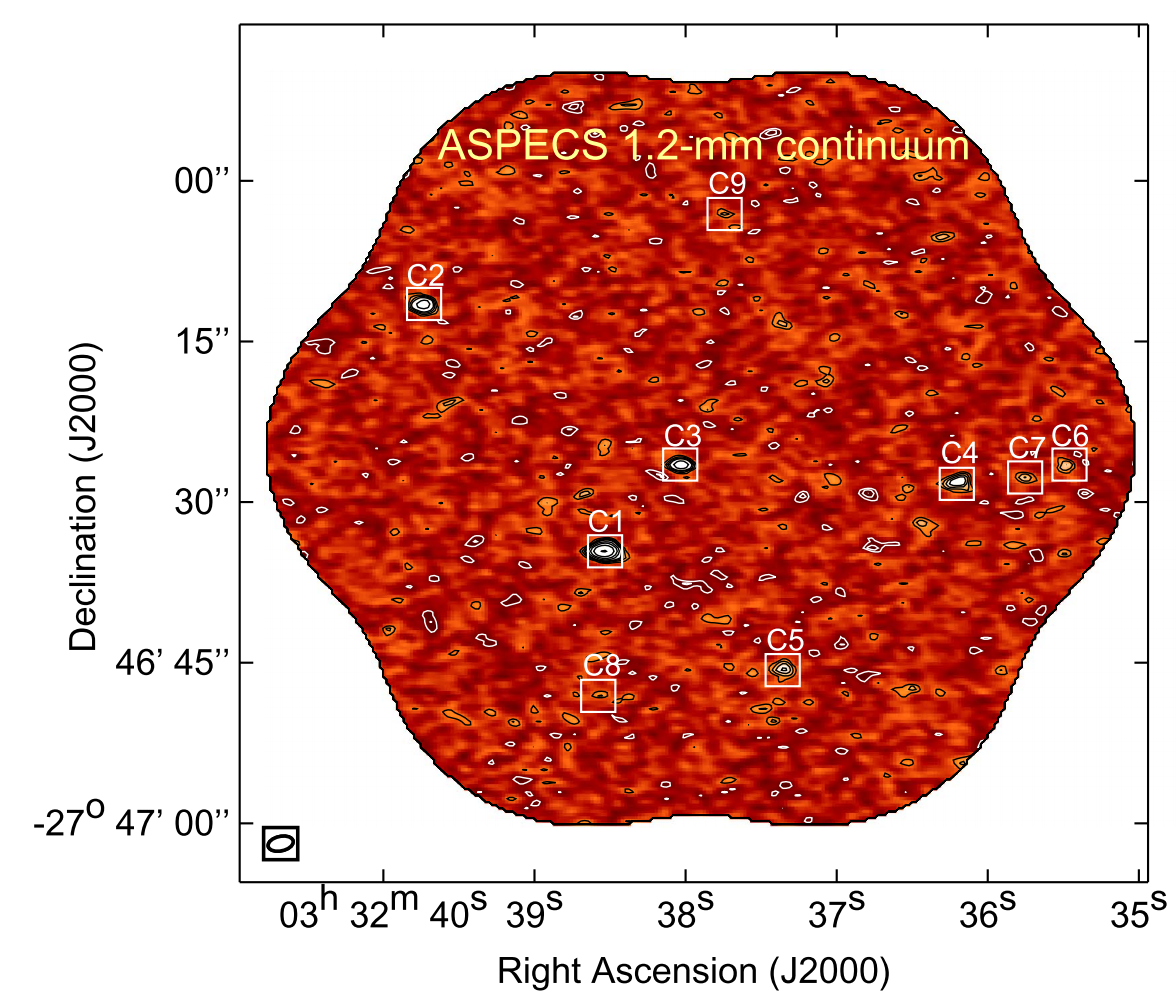}
\caption{ALMA 1.2-mm S/N image obtained by the ASPECS pilot project in the HUDF (from Aravena et al. 2016).Black and white contours show positive and negative emission, respectively.
The boxes show the position of the sources detected with our extraction procedure at S/N$\sim$3.5.}
\label{fig:aspecs_pilot}
\end{figure*}

The ASPECS pilot program (ALMA cycle-2) covered a 1 arcmin$^2$ in the northern part of the HUDF, using a single pointing in band 3 and 7-pointing mosaic
pattern in band 6. The ASPECS pilot region overlaps significantly with an ultra-deep integration obtained with the VLT/MUSE instrument, providing unique spectroscopic coverage over
the redshift range $z=0-6$. The ASPECS pilot 1.2-mm continuum image reached a roughly uniform depth of $\sim13\mu$Jy over this region. The ASPECS pilot program, while covering a
modest area of the sky, probed to be extremely useful to test the techniques necessary and explore que science that could be achieved through these observations. The results from
this survey are presented in a series of seven papers, covering a variety of topics related to galaxy evolution including: the line catalog and survey design (Walter et al. 2016);
the millimeter continuum imaging, number counts and characterization of the faint dusty galaxies (Aravena et al. 2016a); determination of the CO luminosity function and evolution
of the cosmic molecular gas density (Decarli et al. 2016a); characterization of the ISM properties of CO line emitters (Decarli et al. 2016b); [CII] line emitting candidates at
$z=6-8$ (Aravena et al. 2016b); measurements of the infrared excess relation at high redshift (Bouwens et al. 2016); and constraints on CO intensity mapping experiments (Carilli et
al. 2016). Figure \ref{fig:aspecs_pilot} shows the ALMA 1.2-mm map obtained by the ASPECS pilot project. 

\subsection{ASPECS LP: results from ALMA band 3}

The ASPECS large program (LP; ALMA cycles 4 and 5) built on the pilot observations, expanding the covered area to $\sim4.5$ arcmin$^2$, comprising roughly the area known as the
eXtremely Deep Field (XDF) within the HUDF. This region is fully covered by deep VLT/MUSE observations and the deepest HST observations ever obtained. The ASPECS LP thus
complements greatly the legacy value of this field, which will also be covered by Granted Time Observations (GTO) with the James Webb Space Telescope (JWST).

The first results of this survey, based on the ALMA band 3 observations are presented in 5 recent papers (Gonzalez-Lopez et al. 2019; Decarli et al. 2019; Aravena et al. 2019;
Boogard et al. 2019; Popping et al. 2019).

In Gonz\'alez-L\'opez et al. (2019), the full suite of techniques for line detection, cube search, and fidelity and completeness measurements were scrutinized in detail through
comparison of different tools/algorithms, yielding a sample of 16 statistically significant CO line emitters with reliable counterparts. The ASPECS 3mm continuum
map yielded a sample of 6 sources, also detected at 1mm, providing the tightest contraints on the 3mm number counts to date. Interestingly, the identified counterparts to the 3mm
sources seem to lie at significantly high redshifts than the population of 1mm selected sources (e.g. from Aravena et al. 2016).

\begin{figure*}
\centering
\includegraphics[scale=0.4]{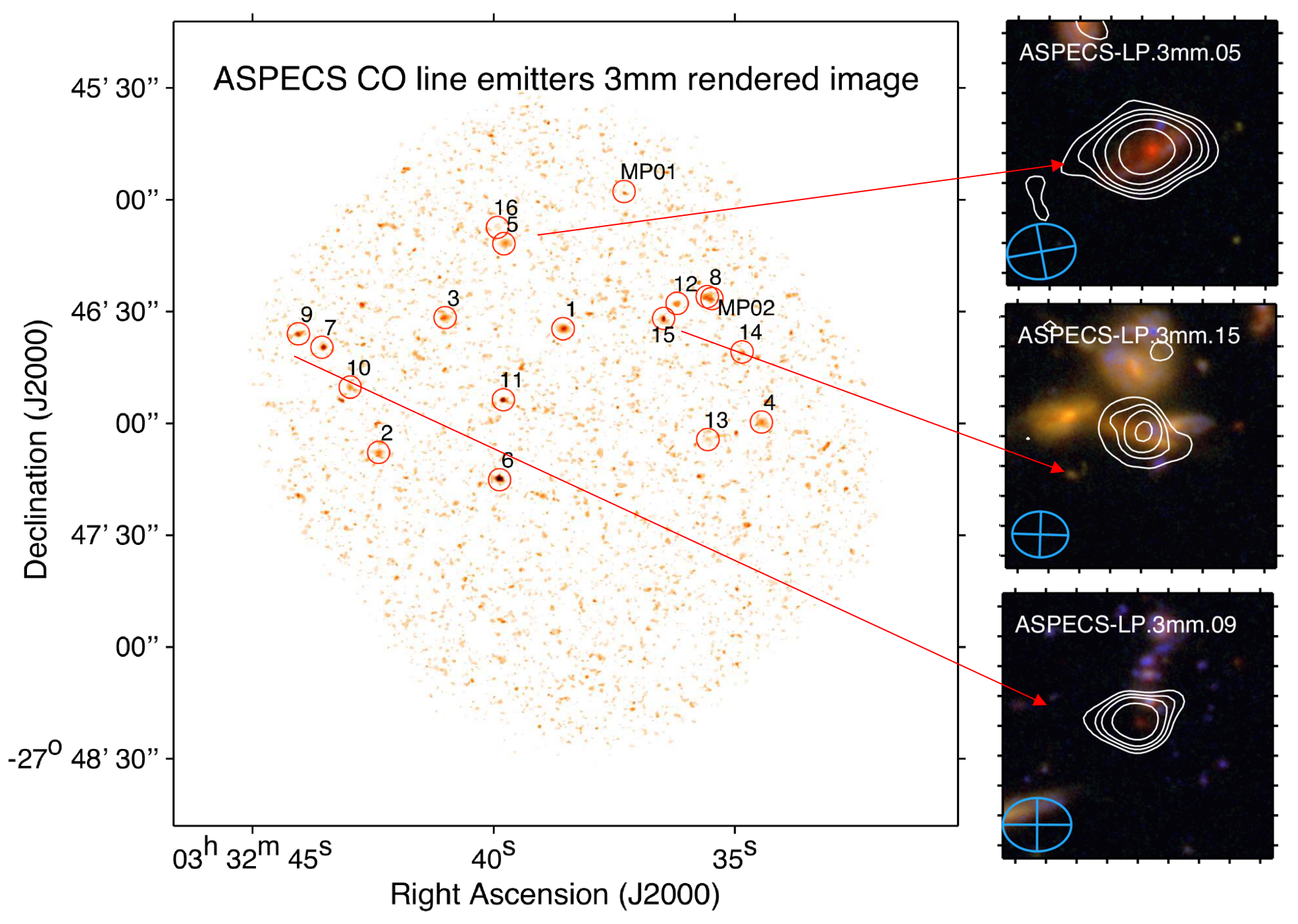}
\caption{{\it Left:)} Rendered CO image obtained by the ASPECS LP toward the HUDF, obtained by co-adding the individual average CO line maps around the bright CO-selected galaxies and the 2 lower significance MUSE-based 
CO sources (adapted from Aravena et al. 2019). ({\it Right:}) HST cutouts with CO contours overlaid on three of the ASPECS sources. These are shown to exemplify the
variety of optical morphologies and brightness of the sources being identified. Source ID9 would not have been pre-selected by targeted follow-up programs.}
\label{fig:corender}
\end{figure*}

The properties of the CO-selected galaxies discovered by the ASPECS survey are presented in Boogaard et al. (2019) 
and Aravena et al. (2019). Results from the former are discussed in a paper in this proceeding series. 
Figure \ref{fig:corender} shows the blank-field CO line detections in the ASPECS LP field, and highlight the optical properties observed in a few examples.
In short, Boogaard et al. (2019) report the search for optical/near-IR counterparts in the HST images, determination of MUSE/VLT spectroscopic redshifts, and 
derivation of physical properties based on spectral energy distribution (SED) fitting using the MAGPHYS code (da Cunha et al. 2008). 
Where available, optical emission lines are used to measure metallicities for the ASPECS galaxies. Using the measured SFRs and stellar masses, Boogaard et al. (2019) study in detail the location of the ASPECS CO-selected galaxies
with respect to the main-sequence of star formation. Furthermore, they find that a significant fraction of the CO selected galaxies
are X-ray sources, suggesting an important presence of AGN activity.  

\begin{figure*}
\centering
\includegraphics[scale=0.5]{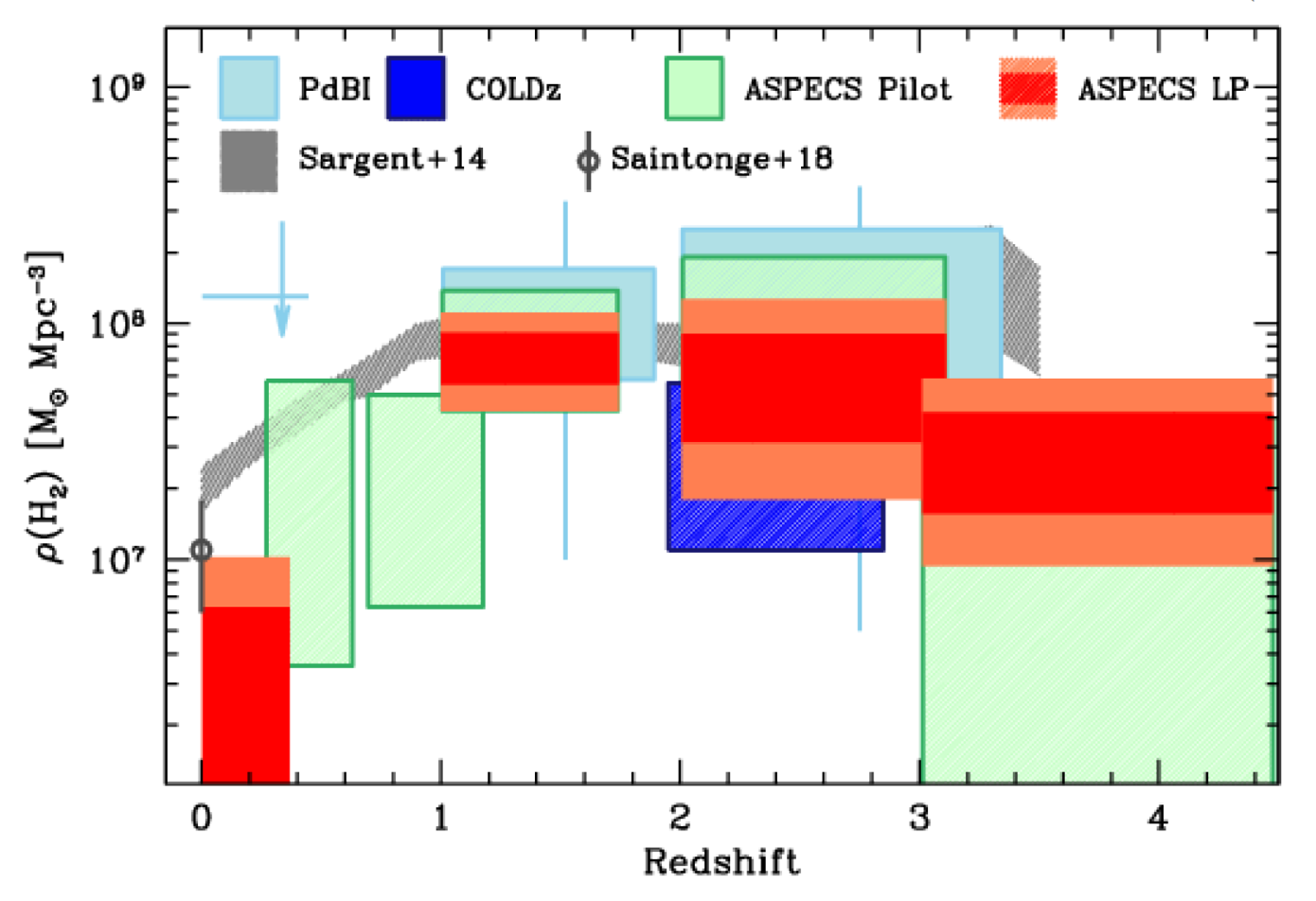}
\caption{Cosmic density of molecular gas (H$_2$) derived by the ASPECS LP survey, compared to previous measurements (from Decarli et al. 2019). 
A rise and fall of the  molecular gas content in galaxies is measured, decreasing by a factor of 6.5 from $z=2.5$ to 0, similar to the evolution of the 
cosmic SFR density.}
\label{fig:cosmicdensity}
\end{figure*}

Through these line search measurements, it is possible to quantify the abundance of CO line emitters as a function of CO 
luminosity. Decarli et al. (2019) provided the key constraints on the CO luminosity function at different redshifts, which can thus 
be converted into measurements of the cosmic molecular gas density as a function of redshift (Fig. \ref{fig:cosmicdensity}). These measurements were found 
to be in excellent agreement with the parallel results from the CO Luminosity Density z (COLDz) survey based on VLA low-J CO 
observations in the COSMOS field (Pavesi et al. 2018; Riechers et al. 2019), and provided the tightest constraints to date on 
the evolution of the molecular gas density. Most importantly, the shape of the cosmic molecular gas density is found to decrease by a factor of $\sim6.5\times$
from $z=1$ to $z\sim0$, in overall consistency with the observed decline in the cosmic SFR density in this redshift range.
This implies that the shape of the cosmic SFR density is likely produced by a similar evolution in the molecular gas reservoirs,
and very likely not due to changes in the star formation efficiencies. These results are found to be in agreement with the
previously observed changes in the molecular gas fractions with cosmic time (e.g. Daddi et al. 2008, 2010; Tacconi et al. 2010, 2013, 2016; 
Saintonge et al. 2013; Schinnerer et al. 2016; Scoville et al. 2017).

Popping et al. (2019) compare the results from the ASPECS CO properties with that of two cosmological galaxy formation models, 
from the IllustrisTNG hidrodynamical simulations and the Santa Cruz semi-analytical model. This study finds that the predicted molecular
gas mass of galaxies at $z>1$ as a function of stellar mass is $2-3\times$ lower than the observations, with the models not being
able to reproduce the number of molecular gas rich galaxies detected with ASPECS. As such the ASPECS observations are found to 
be in tension with these models. 

\begin{figure*}
\centering
\includegraphics[scale=0.33]{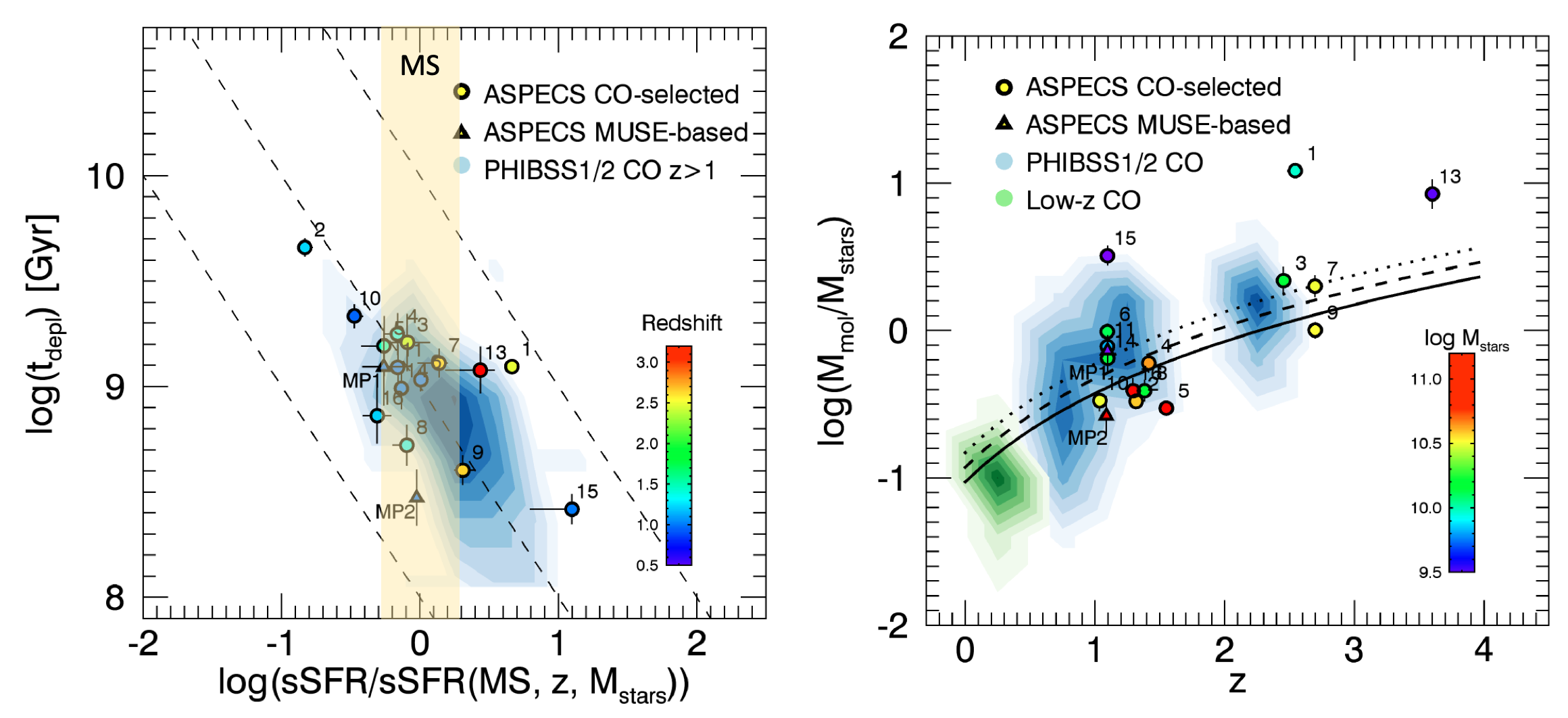}
\caption{{\it (Left:)} Relationship between the molecular gas depletion timescales and normalized sSFR (offset from the MS) for the ASPECS CO galaxies compared to 
literature galaxies at $z>1$ (from Aravena et al. 2019). The location of the MS is highlighted. ASPECS galaxies seem to broadly follow the well established scaling 
relations, yet several galaxies fall above and below the MS. {\it (Right:)} Evolution of the molecular gas fraction with redshift for the ASPECS galaxies compared to 
literature galaxies (from Aravena et al. 2019). The steady decrease of the molecular gas fraction by a factor of roughly an order of magnitude from $z=2$ to 0 is 
well supported by the CO-selected ASPECS sample.}\label{fig:scalingrelations}
\end{figure*}

In complement to this study, Aravena et al. (2019) studied the molecular gas
properties from the CO-selected galaxies. They find that the ASPECS galaxies generally follow 
the ISM scaling relations in terms of the evolution of the molecular gas fraction and gas depletion timescales as 
a function with redshift (Fig. \ref{fig:scalingrelations}). They find that a good fraction of the CO-galaxies are located above and below the MS of star 
formation given the CO-selected nature of the survey. Finally, they find that main sequence galaxies contribute the 
dominant contribution to the cosmic molecular gas density at all redshifts proven, with an increasing/decreasing contribution
from starburst/passive galaxies at increasing redshift (from $z\sim1$ to $z\sim2.5$).

\subsection{ASPECS LP: results from ALMA band 6}

The ALMA band 6 observations reached a uniform depth of $9.3\mu$Jy per beam in the 1.2-mm continuum map, which is unprecedented for an area 
of $\sim4.5$ arcmin$^2$. Many of the results of this part of the ASPECS program, mostly based on the ultra-deep 1.2-mm continuum map, 
have been submitted for publication in Journals as of this writing. 

Gonz\'alez-L\'opez et al. (2019b) finds a sample of 32 significant continuum sources in this area, with additional 26 sources 
found using a optical counterpart priors. Measurements of the number counts as a function of 1.2-mm flux density show that
there is a clear flattening at faint fluxes (below 100$\mu$Jy), which indicates that the observations are able to resolve
most ($\sim90\%$) of the Extragalactic Background Light at 1.2-mm in the HUDF. Interestingly, splitting the number counts in 
bins of stellar mass, SFR, dust mass, and redshift allowed for measurements of the population that dominates the actual aggregate
cosmic dust emission (Gonzalez-Lopez et al. 2019b; Popping et al. 2019b). The comparison of these observed number counts with 
models that use simple assumptions about scaling relations and ISM properties (from Popping et al., 2019b) shows a remarkable 
agreement. Characterization of the physical properties of the dust continuum detected sources shows again a variety, from starburts 
to passive galaxies and confirm previous ASPECS pilot measurements, yielding a median redshift for this population of $z\sim1.5$ (Aravena et al. 2019b).

Using stacking analysis on the 1.2-mm map, Magnelli et al. (2019) measured the evolution of the dust mass function with redshift
for galaxies in the ASPECS footprint. They find that the observations are able to probe deep, below the ``knee'' of the dust mass function. 
By combining the aggregate contribution from different stellar mass ranges, Magnelli et al. are able to recover the evolution of the 
cosmic density of dust mass. After assuming a metallicity-dependent gas-to-mass ratio, they furthermore measure the cosmic density 
of molecular gas. Their results are found to be in good agreement with the CO-based estimates of the cosmic molecular gas density. 

\section{Concluding remarks}

\begin{itemize}
\item In the last 20 years, we have learnt significantly from blank-field deep submm surveys and targeted observations of star-forming galaxies 
in dust and CO line emission. 
\item While these studies have been fundamental to provide a foundational framework for galaxy evolution and the transformation of gas 
to stars in galaxies, they might not be encapsulating the full picture. 
\item A complementary approach is to reach faint SFGs by conducting deep ALMA of surveys 
of dust continuum and molecular line emission in cosmological deep fields, with an increasing number of these surveys to date and on going.
\item Among these surveys, the ASPECS pilot and large programs are proving to provide unique legacy value in the HUDF, in the era of JWST.
\item The observed ASPECS CO-based sources are found to have a diversity of ISM properties, but consistent with “scaling relations”.
\item Molecular gas density measurements based on ASPECS observations are found to be consistent with the evolution of cosmic SFR density.
\item Ultra deep 1mm continuum ASPECS observations are able to place tight constraints on 1-mm number counts, and confirm a flattening of the 1-mm number counts at faint fluxes.
\end{itemize}

\end{document}